\documentclass[reprint,aps,prl,twocolumn,superscriptaddress]{revtex4-1}
\usepackage{graphicx}
\usepackage{mathrsfs}
\usepackage{bm}
\usepackage{hyperref}
\usepackage{dcolumn}
\usepackage{amsmath}
\usepackage{amssymb}
\usepackage{color}

\newcommand{\bohr}{{ \mu_{_B} }}

\newcommand{\bk}{{\mathbf k}}

\begin{document}

\title{Large Anomalous Hall Effect in Topological Insulators \\ Proximitized by Collinear Antiferromagnets}
\author{Chao Lei}
\affiliation{Department of Physics, The University of Texas at Austin, Austin, Texas 78712,USA}
\author{Hua Chen}
\affiliation{Department of Physics, Colorado State University, Fort Collins, CO 80523, USA}
\affiliation{School of Advanced Materials Discovery, Colorado State University, Fort Collins, CO 80523, USA}
\author{Allan H. MacDonald}
\affiliation{Department of Physics, The University of Texas at Austin, Austin, Texas 78712,USA}

\begin{abstract}
CrSb is an attractive material for room-temperature antiferromagnetic spintronic applications 
because of its high N\'{e}el temperature $\sim$700 K
and semi-metallic character.  We study the magnetic properties of CrSb bilayers
on few-layer topological insulator thin films using \emph{ab initio} density functional theory.
We find that the intrinsic parts of the total anomalous Hall conductivities of the thin films are non-zero, and approximately quantized. 
The N\'{e}el temperature of CrSb bilayers on few-layer topological insulator thin films 
is found to be approximately two times larger than that of an isolated CrSb thin film. 
Due to the low Fermi level density of states of CrSb, Hall quantization might 
be achievable by introducing disorder. CrSb bilayers on topological insulator surfaces are therefore 
attractive candidates for high-temperature quantum anomalous Hall effects. 
\end{abstract}

\maketitle

\textit{Introduction---}
The intrinsic contribution to the anomalous Hall effect (AHE) is related to the geometrical phase of Bloch electrons in momentum space \cite{Nagaosa2010,xiao_2010}. When a quasi-two-dimensional anomalous Hall system becomes insulating, its Hall conductivity
is quantized at $e^2/h$ times a topological invariant of the occupied Bloch bands, the 1st Chern number.
The quantum anomalous Hall effect (QAHE) was first predicted in 1988 \cite{haldane_1988}, but 
has been realized experimentally only recently, first in topological insulators (TI) thin films doped with transition metal ions \cite{Yu2010,Chang2013,Checkelsky2014,Chang2015}, then in films of the intrinsic magnetic topological insulator MnBi$_2$Te$_4$ \cite{Otrokov_2017,Li2019_theory,Otrokov2019,Lei2020,Deng_2020,Ge2020,Lei2021_QAH},
and then recently in magic-angle twisted bilayer graphene \cite{Serlin_2019,Sharpe_2019}.
These established QAHE materials exhibit large Hall effects only at relatively low temperatures ($<10$ K),
and are therefore not suitable for the potential applications imagined in low-dissipation spintronics \cite{Smejkal2018}.
High-temperature QAHEs have
been proposed in thin films formed by a TI proximately-coupled to a ferromagnetic insulator with a high Curie temperature ($T_c$), such as EuS, yttrium iron garnet, $\rm Cr_2Ge_2Te_6$ or $\rm Tm_3Fe_5O_{12}$.  Although these films have been realized recently \cite{Katmis2016,Jiang2015_YIG,Che2018_YIG,Mogi_CGT,Tand2017} 
and have high temperature magnetic order with $T_c\approx 300$ K and $>400$ K respectively , the QAHE has not been observed.  


Proximity-induced surface magnetism and anomalous Hall effects 
can also be achieved by establishing atomically sharp interfaces between a TI and an A-type (consisting of ferromagnetic layers that alternate in orientation) antiferromagnet (AFM) \cite{He2016,He2017}.
The AFM can locally magnetize the surface of the TI through short-range interfacial exchange 
coupling.  In Refs.~\onlinecite{He2016,He2017} $\rm CrSb$ was used as the AFM, motivated by the material's high 
bulk N\'{e}el temperature ($T_N$) of around 700 K\cite{Willis1953,Snow1953,Takei1963}. 
In this paper we study the magnetic properties and the intrinsic anomalous Hall conductivity (AHC) of CrSb/TI heterojunctions 
using \emph{ab initio} density functional theory (DFT). We show that the AHC of a single bilayer (defined below) of CrSb is 
not only non-zero, as expected because the bilayer does not possess the bulk's invariance under a combination 
of time-reversal and translation, but also large - around $0.64~e^2/h$.  

The AHE is traditionally associated with ferromagnetism.  Its presence at a AFM/TI heterojunction is reminiscent of recent 
discoveries of AHE's in some bulk antiferromagnets, including noncollinear antiferromagnets \cite{Chen2014,Nakatsuji2015}, and 
also collinear antiferromagnets \cite{smejkal2020crystal,Li2019,Feng2021} that do not possess time-reversal-like symmetries forbidding the AHE.
We find \cite{Experiment} that the AHC increases to $1.2~e^2/h$ when bilayer CrSb is placed on a 3 quintuple layer (QL) Bi$_2$Te$_3$ TI film, 
and further to $4~e^2/h$ if 5 QL of Bi$_2$Te$_3$ are used.  
Moreover, our DFT results suggest that exchange coupling in bilayer CrSb on TI thin films is twice as strong as in  
isolated bilayers.  Exchange enhancement has also been found experimentally in some
ferromagnetic-insulator/TI heterojunctions \cite{Katmis2016}, and can
potentially lead to high N\'{e}el temperatures.  For CrSb bilayers on a TI we estimate using mean-field theory that 
the N\'{e}el temperature is about 1200 K, nearly twice the $T_N$ of bulk CrSb. 
Since bulk CrSb is a bad metal \cite{Suzuoka1957} with a very small Fermi surface, it
might be experimentally feasible to open a mobility gap in intentionally disordered devices, and
in this way to achieve perfect Hall quantization at high temperatures.



\textit{Magnetic properties---}
Bulk CrSb is an A-type AFM 
with local moments on the Cr ions and a nickel-arsenide (NiAs) structure \cite{Snow1953}. 
The Cr atoms form triangular lattices with AA stacking along the $c$ axis (see supplemental material for details \cite{supplement}).
When magnetically ordered, each (0001) layer of CrSb contains two Cr atomic layers, and is therefore referred to here 
as a single bilayer. 
In our DFT calculations, we use the experimental lattice constants, 
$a= 4.121 {\rm \AA}$ \cite{Takei1963} for the nearest-neighbor Cr-Cr distance in a (0001) plane
and $c=5.47 {\rm \AA}$, which is twice the distance between Cr layers.
A single CrSb bilayer is terminated by Cr and Sb atoms respectively at its two surfaces \cite{supplement}.
The two Cr atoms in each 2D unit cell are therefore distinguishable, 
which implies that the total magnetization will be non-zero in spite of overall AFM order in bulk. 
(We expect that this 2D ferrimagnetism will persist for thin CrSb films with only a few bilayers.) 
As shown in Table~\ref{mag_moment}, the net magnetization of an isolated CrSb bilayer
is about 0.7 $\bohr$ per 2D unit cell according to the DFT calculations. 
The moments decrease in magnitude in both layers when the CrSb bilayer is placed on top of $\rm Bi_2 Te_3$,
but the difference between layers is maintained, and converges as the TI thickness is varied.  
For comparison we also show results in Table~\ref{mag_moment} for the ferromagnetic configuration of the 
moments, which has a higher energy, and for the case of bulk CrSb in 
which the two Cr atoms in each unit cell are indistinguishable and have the same local moment magnitude.
In all cases the local moment magnitude is $\sim$3-4 $\bohr$.  

\begin{table}[h]
\caption{\label{mag_moment}
Local magnetic moments (in $\bohr$) of Cr atoms in bulk CrSb, in a CrSb bilayer without a TI substrate, 
and in a CrSb bilayer on multiple quintuple layers of $\rm Bi_2 Te_3$. $N$ denotes the number of TI quintuple layers.
}
\begin{ruledtabular}
\begin{tabular}{c c | c c c c c }
 CrSb/(TI)$_N$& Bulk & $N$ = 0 & $N$ = 3 & $N$ = 4 & $N$ = 5 & $N$ = 6 \\ [0.5ex]
 \hline
 Cr$_1$(AF) & 2.977 & 3.954 & 3.424 & 3.444 & 3.445 & 3.441 \\
 Cr$_2$(AF) & -2.977 & -3.282  & -2.960 & -2.938 & -2.937 & -2.935 \\
 Total & 0  & 0.672 & 0.464 & 0.506 & 0.506 & 0.506 \\
 \hline
 Cr$_1$(FM)& 2.956 & 4.049 & 3.392 & 3.424 & 3.400 & 3.417 \\
 Cr$_2$(FM) & 2.956 & 3.222 & 2.683 & 2.666 & 2.673 & 2.654 \\
 Difference & 0 & 0.827  & 0.709 & 0.758 & 0.727 & 0.763 \\
 \end{tabular}
 \end{ruledtabular}
\end{table}

\begin{figure}
  \centering
  \includegraphics[width=0.9 \columnwidth]{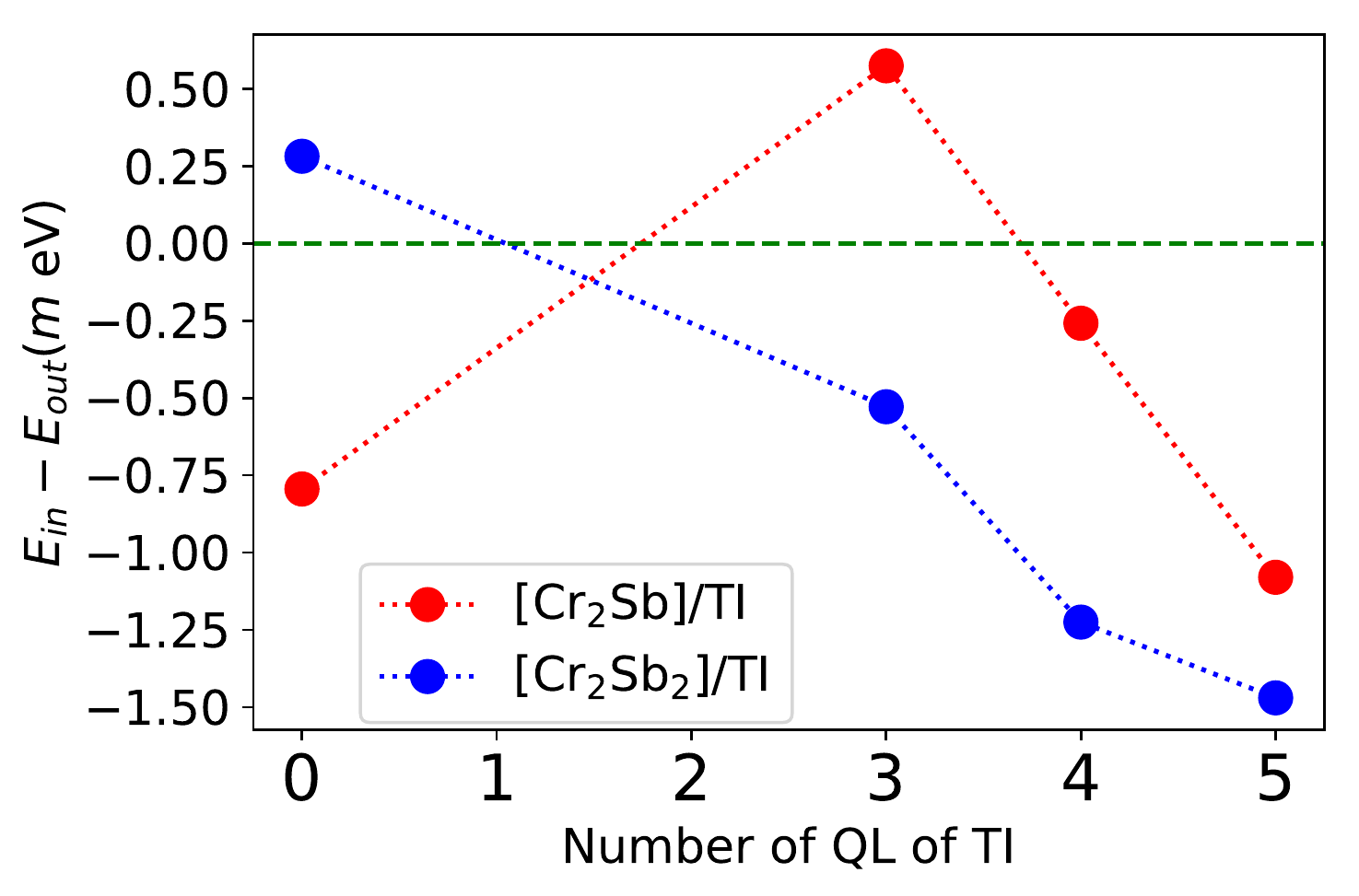}
  \caption{Magnetic anisotropic energy for a CrSb bilayer on $\rm Bi_2 Te_3$ substrates with different thicknesses. $E_{in}$ is the ground state energy per Cr atom when the Cr moments are in-plane, while $E_{out}$ is the energy
  when the Cr moments are out-of-plane. As shown in the supplemental material~\cite{supplement}, the easy directions are always either purely in-plane or purely out-of-plane. The green dotted line labels the boundary between in-plane and out-of-plane anisotropy. 
  Above the line the easy direction is out-of-plane, while below the line the easy direction is in-plane. 
  The red and blue curves are for the case of a CrSb bilayer and the case of a Cr$_2$Sb layer 
  , in which one Sb atom is removed from the bilayer. }\label{crsb_mae}
\end{figure}

In spite of the net magnetization, we find that in CrSb bilayer the local moments prefer an out-of-plane orientation as in bulk CrSb\cite{Snow1953}.
As shown in Fig.~\ref{crsb_mae}, the magnetic anisotropy energy for a CrSb bilayer without the TI substrate is around 0.25 meV per Cr atom. However, when the CrSb bilayer is placed on the TI substrate, the easy axis of the CrSb film is in-plane and the anisotropy energy 
increases in magnitude, ranging from 0.5 meV (3 QL) to 1.5 meV (5 QL) per Cr atom. 
This large magnetic anisotropy energy decreases the relevance of 2D magnetiztion-direction fluctuations, which 
are otherwise suppressing the order temperatures in quasi-2D magnets. 
Magnetic anisotropy is sensitive to the coordination of the two Cr atoms, as illustrated by the red curves in 
Fig.~\ref{crsb_mae} calculated with one Sb layer removed from the bilayer. 
In this case the easy axis of CrSb on a 3-QL TI is out-of-plane, whereas a freestanding Cr$_2$Sb film has an in-plane easy axis. 
We consider only the case of out-of-plane magnetic moments in the following anomalous Hall conductivity calculations. 

In addition to the magnetic anisotropy energy, the magnetic ordering temperature is also important for applications of 
magnetized TI surface states. The low Curie temperature of a magnetically doped TI limits its potential for applications. 
In the past researchers have established \cite{Katmis2016} that interfaces between a ferromagnetic insulator and a TI 
can exceed the critical temperature of the isolated ferromagnet, possibly 
because of strong spin-orbit coupling contributed by the TI \cite{Kim2017}.
We have calculated the magnetic exchange interactions in CrSb bilayers by mapping the energies of several configurations 
that are meta-stable in DFT to classical Cr-spin Hamiltonians, with interlayer nearest-neighbor exchange coupling $J_v$ and intralayer nearest-neighbor exchange coupling $J_1$ (The details are provided in the supplemental materials \cite{supplement}). 
The truncation of the spin-Hamiltonian is motivated by bulk CrSb calculations  
that the second nearest neighbor in-plane exchange coupling ($J_2 = 0.29$ meV) is 
much smaller than the nearest exchange coupling ($J_1 = 7.62$ meV). The values of the exchange couplings obtained in this way are summarized in Table~\ref{exchange_constant}, where we find that the interlayer and intralayer exchange couplings in CrSb bilayer are similar to 
those in bulk CrSb.  The interlayer exchange coupling $J_v$ in bulk CrSb is around -17.83 meV, 
{\it vs.} -19.14 meV for CrSb bilayer, and roughly doubles in magnitude the intralayer exchange coupling which is 7.62 meV,
{\it vs.} 8.76 meV for CrSb bilayers. However, when the CrSb bilayer is placed on the top of the TI substrate, both $J_v$ and $J_1$ are significantly enhanced, by approximately a factor of two.

\begin{table}[h]
\caption{\label{exchange_constant}
Magnetic exchange interactions between Cr atoms obtained via the method detailed in supplemental material \cite{supplement}. 
$J_v$ is the interlayer nearest-neighbor exchange coupling, and $J_1$ is the intralayer coupling constant.}
\begin{ruledtabular}
\begin{tabular}{c c c}
 Material & $J_v$ (meV) & $J_1$ (meV)  \\ [0.5ex]
 \hline
 Bulk CrSb & -17.83 & 7.62  \\
 Isolated CrSb bilayer & -19.14 & 8.76  \\
 CrSb bilayer/3 QL TI & -35.22 & 17.68  \\
 CrSb bilayer/4 QL TI & -35.74 & 17.55 \\
 CrSb bilayer/5 QL TI & -35.79 & 17.64  \\
 CrSb bilayer/6 QL TI & -35.86 & 17.52  \\
 \end{tabular}
 \end{ruledtabular}
\end{table}

Using the values of the exchange couplings in Table~\ref{exchange_constant}, we estimated the N\'{e}el temperatures using mean-field theory.
(Details are provided in the supplemental material \cite{supplement}.)  In Fig.~\ref{neel_temperature} we plot the  
quintuple-layer number ($N$) dependence of $T_{MF}$, which shows that the CrSb bilayer has a slightly lower $T_{MF}$ than 
bulk due to the smaller coordination number of each Cr atom, even though the exchange couplings in the CrSb bilayer are slightly 
larger.  (See Table~\ref{exchange_constant}.) However, when the CrSb bilayer is placed on a $\rm Bi_2 Te_3$ thin film, $T_{MF}$ is much larger than that in bulk CrSb, and up to around 2050 K. The N\'{e}el temperature estimated from mean-field theory is of course 
larger than the expected value.  In bulk CrSb $T_{MF} \approx 1150 K$, 
which implies a ratio of $T_{N}/T_{MF} \approx 0.6$ when the experimental value ($\sim 700$ K) is used
for the Ne\'el temperature.  In comparison, the ratio $T_{c}/T_{MF} \approx 0.56$ in the case of a 2D Ising model.
Quantitatively the ratio of the actual critical temperature to the mean-field critical temperature is expected to 
increase with space-dimension and with magnetic anisotropy strength. 
If we assume that we are close to the strong anisotropy Ising limit, the N\'{e}el temperatures of the CrSb bilayers on TI substrates
would be about 1200 K, which is much higher than room temperature and also higher than the N\'{e}el temperature of bulk CrSb.

\begin{figure}[h]
  \centering
  \includegraphics[width=0.9 \columnwidth]{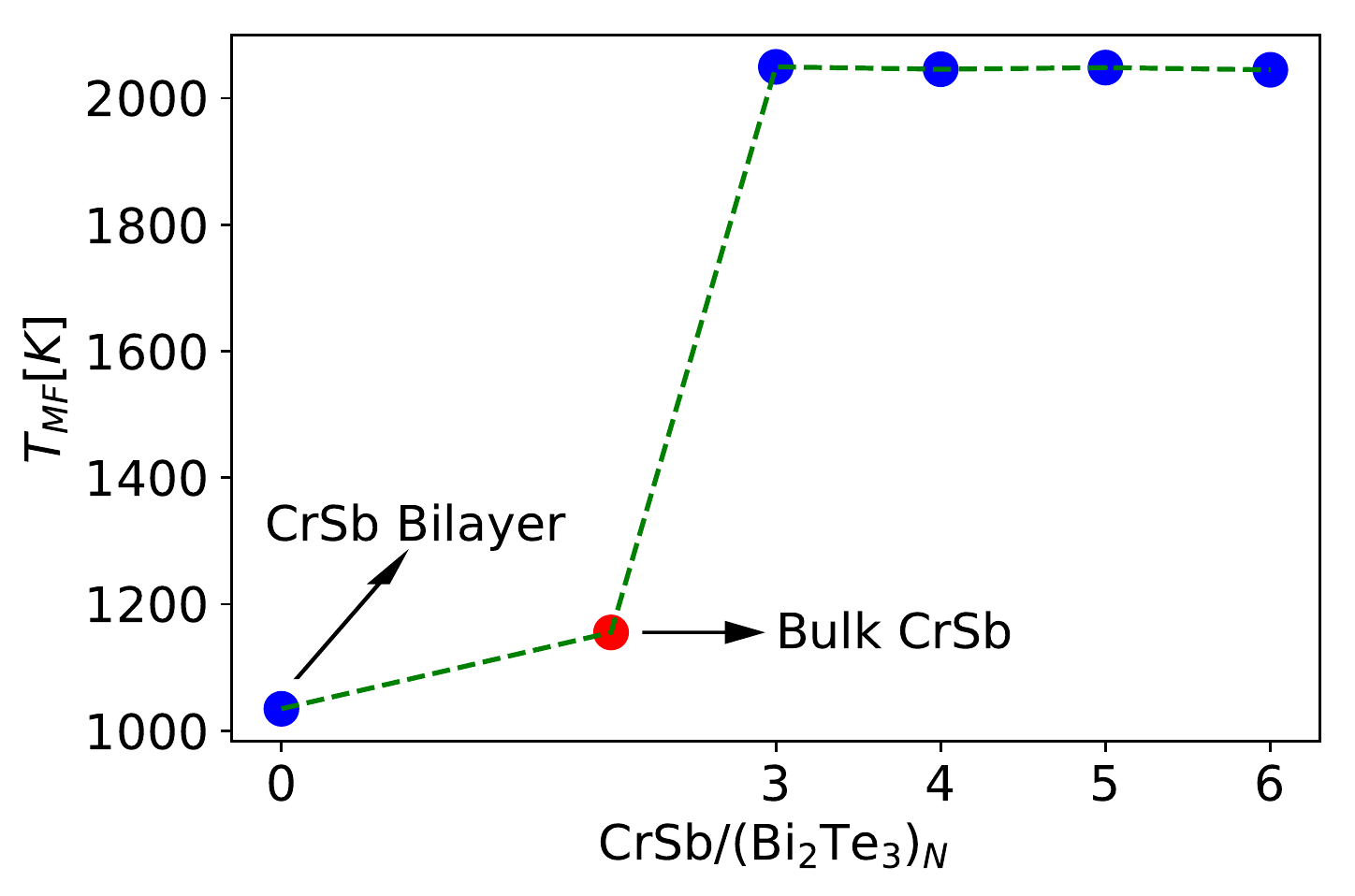}
  \caption{Layer dependence of N\'{e}el temperature from mean-field theory. The red dot stands stands for bulk CrSb. The horizontal axis is the number of QLs of $\rm Bi_2 Te_3$, with $N=0$ the case of freestanding CrSb bilayer. 
  }\label{neel_temperature}
\end{figure}

\textit{Anomalous Hall Conductivity---}
Building on the pioneering work by Karplus and Luttinger, Smit, and Berger \cite{Nagaosa2010}, it is now widely 
accepted that there are three main mechanisms contributing to the anomalous Hall effect: the intrinsic contribution due to electric-field-induced inter-band coherence in a perfect crystal \cite{Karplus1954}, skew scattering \cite{smit1955,smit1958}, and side-jump scattering \cite{berger1970side}. Moreover, it has become more clear recently that the Karplus-Luttinger mechanism is a manifestation of momentum-space Berry curvature of Bloch states \cite{xiao_2010}. The latter, whose integral over 2D Brillouin zone is the 1st Chern number times $2\pi$, is the sole contribution to the Hall conductivity in the case of quasi-two-dimensional insulators,
which can therefore exhibit a quantized Hall conductivity.  Since the CrSb/TI thin films we study are normally 
two-dimensional semimetals, we consider only the 
Karplus-Luttinger mechanism. To this end we first study the electronic structures of CrSb bilayer. 

Transport experiments have shown that bulk CrSb behaves as a semiconductor below the N\'{e}el temperature \cite{Suzuoka1957} and as a metal above it, although previous DFT calculations find that CrSb is a semimetal \cite{kahal_2007}. 
The resistivity $\rho$ for bulk CrSb at room temperature is around $\rm 3 \times 10^{-4}~\Omega cm$, which is several times larger than that of the classic semi-metal bismuth ($\rho \simeq \rm 1.3 \times 10^{-4}~\Omega cm$).

\begin{figure}[h]
  \centering
  \includegraphics[width=0.9 \columnwidth]{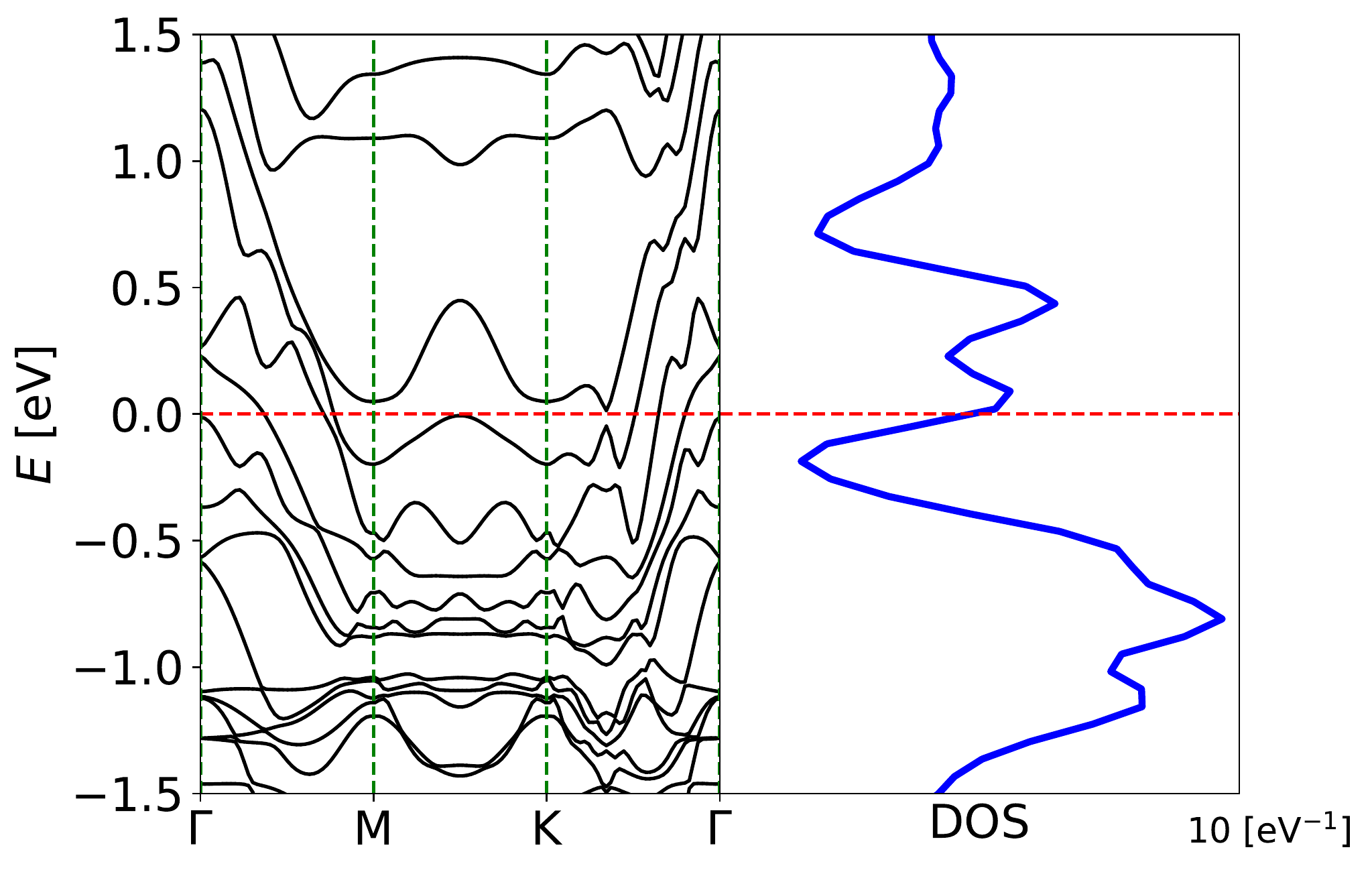}
  \caption{Bandstructure (left) and density of sates (DOS, right) of an isolated CrSb bilayer. 
  The left-hand side of the DOS panel corresponds to DOS$=0$}\label{crsb_band}
\end{figure}

The density of states (DOS) of a freestanding CrSb bilayer in the AFM state has a minimum close to the Fermi energy
(shown in Fig.~\ref{crsb_band}), as does bulk CrSb.  
The Kohn-Sham band structure plot shows four subbands crossing the Fermi level, which are mostly derived from the $3d$ orbitals of Cr atoms. 
For the case of CrSb/$\rm Bi_2 Te_3$ heterojunctions two of the four subbands are pushed below the Fermi level (as shown in the supplemental materials \cite{supplement}), demonstrating strong hybridization between
the Cr $d$ orbitals and the surface states of $\rm Bi_2 Te_3$, leaving two subbands 
that cross the Fermi level near the $\Gamma$ point and derive mainly from $\rm Bi_2 Te_3$ orbitals.
The Fermi level DOS is smaller in the CrSb/TI case.  


\begin{figure*}[htp]
  \centering
  \includegraphics[width=2.0 \columnwidth]{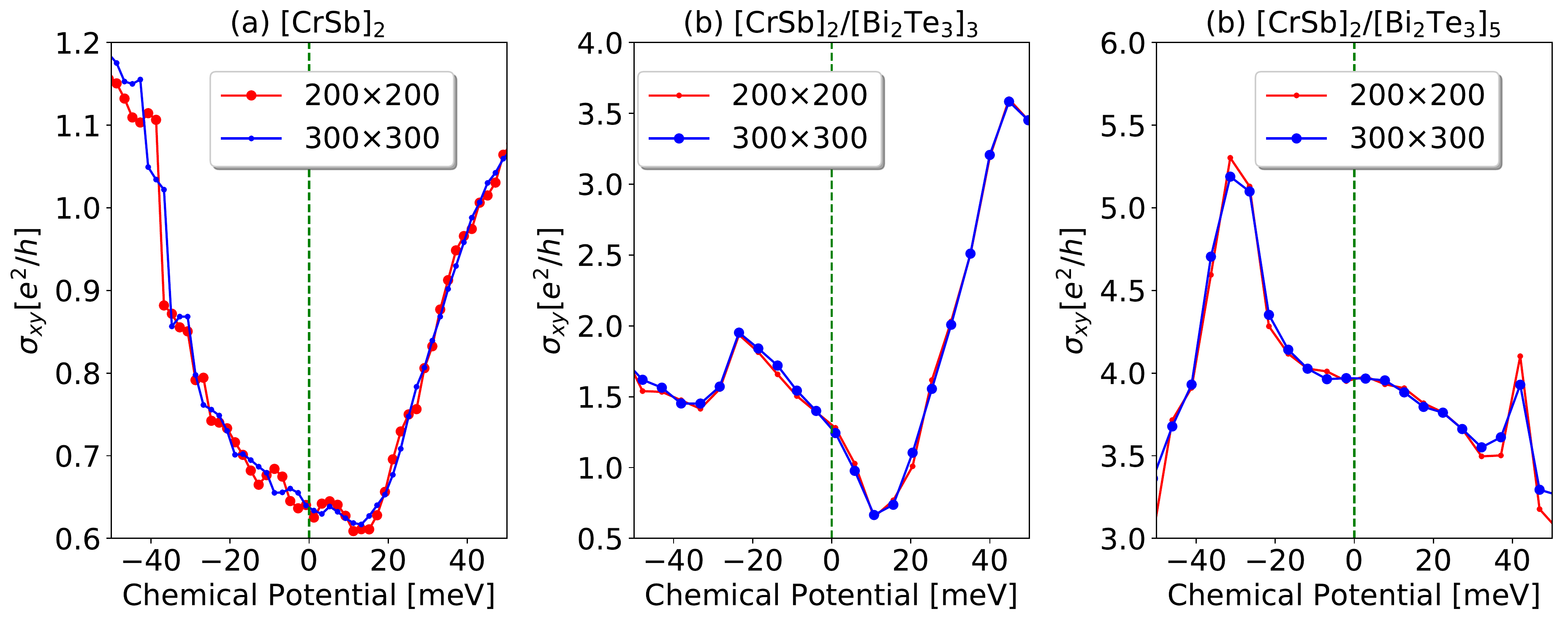}
  \caption{Anomalous Hall conductivity {\it vs.} chemical potential for a (a) CrSb bilayer, (b) a CrSb bilayer on 3 QL $\rm Bi_2 Te_3$, and (c) a CrSb bilayer on 5 QL $\rm Bi_2 Te_3$. Integration over the Brillouin zone was carried out using $200\times200\times1$ ($5\times5\times1$ for adaptive refinement) and $300\times300\times1$ ($7\times7\times1$ for adaptive refinement) $\bk$-meshes, with the results labeled separately in each plot.
  }\label{ahc_theory}
\end{figure*}

Given the Kohn-Sham bands, we are able to calculate
the intrinsic anomalous Hall conductivities of all thin film structures
using the Wannier interpolation technique \cite{Wang2006,supplement}. 
Fig.~\ref{ahc_theory} presents intrinsic anomalous Hall conductivities 
for a free-standing CrSb bilayer and for CrSb/TI heterojunctions {\it vs.} chemical potential near the Fermi level.
The Hall conductivities are large in all cases, and tend to be larger for thicker TI films.
In the free-standing CrSb bilayer case, as illustrated in Fig.~\ref{ahc_theory} (a), 
the anomalous Hall conductivity has a minimum near the Fermi level. 
In Fig.~\ref{ahc_theory} (b) and (c), $\sigma_{xy}$ changes relatively slowly with chemical potential near Fermi level.
The Hall conductivities of CrSb/TI heterojunctions have values close to $e^2/h$ and $4e^2/h$ for 3 and 5 quintuple layers TI respectively. 
The relatively small deviation from quantized value suggests
that it may be possible to realize a quantized anomalous Hall effect if a mobility gap can 
be induced by increasing disorder. 

Large intrinsic anomalous Hall conductivities are often associated with hot-spots in momentum space 
that host weakly split band states that span the Fermi energy.  
To identify the dominant contributions to the Brillouin zone integrals
we calculated the Berry curvature along the high-symmetry lines together with the band structure \cite{supplement}. 
For a free-standing CrSb bilayer the main contribution to Berry curvature 
comes from the momentum near the K point, while for the case of CrSb/TI heterojunctions the major contribution comes from regions near the $\Gamma$ point. 
Even though the 3QL CrSb/TI heterojunction has a Berry curvature between the $\Gamma$ and K points that
is one order of magnitude larger than that in the case of 5 QL $\rm Bi_2 Te_3$, the integrated AHC 
is larger in the 5QL case.  (See Fig.~\ref{ahc_theory}.)  The Berry curvature hot-spots are not necessarily on 
high-symmetry lines and are extremely sensitive to electronic structure details.

\begin{table}[htp]
\caption{\label{ahc_exp}
Recent experimental values of the anomalous Hall resistivity (AHR) involving CrSb and TI heterostructures \cite{He2016,He2017}. Note that $h/e^2\approx 25.812$ $\rm k \Omega$.
}
\begin{ruledtabular}
\begin{tabular}{c c }
 Heterojunctions & AHR ($\rm k\Omega$) \\
 CrSb(22 bilayers) & $\thicksim 10^{-3} $ \\
 CrSb(22 bilayers)/TI(3 QL) & $0.01\hbox{-}0.1 $ \\
 CrSb(22 bilayers)/TI(3 QL)/CrSb(22 bilayers) & $\thicksim 0.2 $ \\
 CrSb(13 bilayers)/MTI\footnote{Cr-doped (Bi,Sb)$_2$Te$_3$}(3QL) & $\thicksim 0.8 $ \\
 $\rm [CrSb(13 bilayers)/MTI(3QL)]_4$ & $\thicksim 0.6 $ \\
 MTI(9QL) & $\thicksim 0.2 $ \\
 $\rm [MTI(9QL)/CrSb(44 bilayers)/MTI(9QL)]_x$ & $\thicksim 0.6\hbox{-}0.8 $ \\
 \end{tabular}
 \end{ruledtabular}

\end{table}
\textit{Discussion---}
In summary, motivated by the large bulk N\'{e}el temperature $\sim $ 700 K of bulk CrSb,
we have studied the magnetic properties of free-standing CrSb bilayers and CrSb bilayers on 
thin TI films.  We find that magnetic exchange interactions strengthen substantially when the bilayers are placed on 
TIs, and estimate critical temperatures that are easily in excess of room temperature.  
Because the inversion symmetry of bulk CrSb is broken in bilayers, the thin films are 
ferrimagnetic with a net magnetization around 0.7 $\bohr$ per unit cell.
The same reduction in symmetry leads to large anomalous Hall effects.  We find that the anomalous Hall effect strengthens
on TI substrates \cite{He2016,He2017}. Since that CrSb is a semimetal with a small Fermi surface, a bulk mobility gap at the Fermi level 
can potentially be induced opened by disorder.  If so the large intrinsic AHC may survive 
and would then necessarily be quantized. 

Experimental result from Refs.~\onlinecite{He2016, He2017} on the 2D anomalous Hall resistance (AHR) of CrSb/TI heterojunctions and CrSb/magnetized TI heterojunctions are summarized in Table~\ref{ahc_exp}.
Many of these observations were made on samples in which the TI has also been made magnetic by Cr doping.
Generally speaking, these observations find Hall conductivities that are much smaller than in our calculations.
Although this difference could in principle be related to interface quality, we point out that all observations were made on
thin films with many layers of CrSb.  As we have emphasized the difference between the environments of antiferromagnetically coupled 
Cr sites is much larger in the CrSb bilayer case.  Our calculations therefore motivate anomalous studies of 
thin films in which a single CrSb bilayer is grown on TIs.

\section*{Acknowledgments}  This work was financially supported by the Army Research Office under Grant Number W911NF-16-1-0472. 
HC was partially supported by NSF CAREER grant DMR-1945023. The authors acknowledge the Texas Advanced Computing Center (TACC) at The University of Texas at Austin for providing HPC resources that have contributed to the results in this paper.


\bibliography{TICrSb}

\end{document}